\documentclass[5p]{elsarticle} 

\usepackage{atlasphysics}
\usepackage{subfigure}
\usepackage{xspace}
\usepackage{lineno}
\usepackage{multirow}

\usepackage{units}
\usepackage{hyperref}

\def\Ztotautau{\ensuremath{Z \to \tau\tau}\xspace}

\def\MET{\ensuremath{E_{\mathrm{T}}^{\mathrm{miss}}}\xspace}
\def\METx{\ensuremath{E_{\mathrm{x}}^{\mathrm{miss}}}\,}
\def\METy{\ensuremath{E_{\mathrm{y}}^{\mathrm{miss}}}\,}

\def\pt{\ensuremath{p_{\mathrm{T}}}\xspace}

\def\ttbar{\ensuremath{t\bar{t}}\xspace}

\newcommand{\lumi}{34~pb$^{-1}$}

\newcommand{\xsec}{7.2}
\newcommand{\xsecstat}{0.2}
\newcommand{\xsecsyst}{1.1}
\newcommand{\xseclumi}{0.2}
\newcommand{\xsectot}{11.1}
\newcommand{\xsectotstat}{0.3}
\newcommand{\xsectotsys}{1.7}
\newcommand{\xsectotlumi}{0.4}

\newcommand{\xsecfid}{0.70}
\newcommand{\xsecfidstat}{0.02}
\newcommand{\xsecfidsyst}{0.11}
\newcommand{\xsecfidlumi}{0.02}

\newcommand{\NA}{2335~}
\newcommand{\NAQCDal}{127~}

\newcommand{\NAQCDsys}{9~}

\newcommand{\NAEW}{284~}

\newcommand{\NAEWsys}{43~} 
\newcommand{\AW}{0.0975~}
\newcommand{\AWstat}{0.0004~}
\newcommand{\AWsys}{0.0019~}
\newcommand{\CW}{0.0799~}
\newcommand{\CWstat}{0.0011~}
\newcommand{\CWsys}{0.0107~} 

\def\MET{\ensuremath{E_{\mathrm T}^{\mathrm{miss}}}\xspace}
\def\METSIG{\ensuremath{S_{E_{\mathrm T}^{\mathrm{miss}}}}\xspace}
\def\SumET{\ensuremath{\sum{\ET}}\xspace}

\def\pT{\ensuremath{p_{\mathrm T}}\xspace}

\def\tauhadpT{\ensuremath{p_{\mathrm T}^{\mathrm{\tauhad}}}\xspace}

\def\tauhad{\ensuremath{\tau_{\mathrm h}}\xspace}

\def\Wtaunu{\ensuremath{W\to\tau_{\mathrm{h}}\nu_\tau}\xspace}

\def\Wenu{\ensuremath{W\to e\nu_e}\xspace}
\def\Wmunu{\ensuremath{W\to\mu\nu_\mu}\xspace}
\def\Zee{\ensuremath{Z\to ee}\xspace}

\def\Ztautau{\ensuremath{Z\to\tau\tau}\xspace}
\def\ttbar{\ensuremath{t\bar{t}}\xspace}

\def\sigmafid{\ensuremath{\sigma^{\mathrm{fid}}_{W\rightarrow\tauhad\nu_\tau}}}
\def\sigmatot{\ensuremath{\sigma^{\mathrm{tot}}_{W\rightarrow\tauhad\nu_\tau}}}
\def\sigmatotfullw{\ensuremath{\sigma^{\mathrm{tot}}_{W\rightarrow\tau\nu_\tau}}}

\def\ipb{pb$^{-1}$\xspace}

\def\NEW{\ensuremath{N_{\mathrm{EW}}}\xspace}
\def\NQCD{\ensuremath{N_{\mathrm{QCD}}}\xspace}
\def\EW{EW}

\journal{Physics Letters B}

\begin{document}
\begin{frontmatter}
\title{\vspace{-0.5cm} \flushright{\normalsize{CERN-PH-EP-2011-122}} \\ \center{
Measurement of the $W\rightarrow\tau\nu_\tau$ Cross Section in $pp$ Collisions at \mbox{$\sqrt{s}$ = 7~TeV} with the ATLAS experiment}
}

  \author{The ATLAS Collaboration}
  \address{}

\begin{abstract}
The cross section for the production of  $W$ bosons with subsequent decay $W\to\tau\nu_\tau$ 
is measured with the ATLAS detector at the LHC. The analysis is based on a data sample that was recorded in 2010 at a proton-proton center-of-mass energy of $\sqrt{s}=7\TeV$ and corresponds to an 
integrated luminosity of \lumi. The cross section is measured in a region of high detector acceptance and then extrapolated to the full phase space. The product of the total $W$ production 
cross section and the $W\to\tau\nu_\tau$ branching ratio is measured to be 
$\sigmatotfullw = \xsectot \pm \xsectotstat \mathrm{~(stat)} \pm \xsectotsys \mathrm{~(syst)} \pm \xsectotlumi \mathrm{~(lumi)} \mathrm{~nb}$.
\end{abstract}

\begin{keyword}
W boson, Standard Model, tau lepton
\end{keyword}

\end{frontmatter}

\section{Introduction}
\label{sec:introduction}
The study of processes with $\tau$ leptons in the final state is an important part of the ATLAS physics program, for example in view of searches for the Higgs boson
or supersymmetry~\cite{AtlasCSCBook,paper:AHtautau,note:tHplus}. 
Decays of Standard Model particles to $\tau$ leptons, in particular \mbox{$Z\to\tau\tau$} and $W\to\tau\nu_\tau$, are important background processes in such searches.
Studies of the $W\to\tau\nu_\tau$ decay complement the measurement of W production in the muon and electron decay modes~\cite{ATLAS-CONF-2011-041,WZnewpaper}.
In addition, $W\to\tau\nu_\tau$ decays can be used to validate the
reconstruction and identification techniques for $\tau$ leptons and
the measurement of the missing transverse energy (\MET), which are
both fundamental signatures in a wide spectrum of measurements at the LHC.

At next-to-next-to-leading order (NNLO), the \mbox{$W\to\tau\nu_\tau$} signal is predicted to be produced at $\sqrt{s}=7\TeV$ with a cross section times branching ratio  of $\sigma\times\mbox{BR} = 10.46 \pm 0.52 \mathrm{~nb}$~\cite{ref:FEZW,ref:MSTW2008,Aad:2010yt}.
Since purely leptonic $\tau$ decays 
cannot be easily distinguished from electrons and muons from $W\to e\nu_e$ or \mbox{$W\to\mu\nu_\mu$} decays, the analysis presented in this paper uses only hadronically decaying $\tau$ leptons (\tauhad). 
Events from \mbox{$W\to\tau\nu_\tau$} production contain predominantly
low-\pt $W$ bosons decaying into $\tau$ leptons with typical visible transverse momenta between 10 and 40~\GeV.
In addition, the distribution of the missing transverse energy, associated with the neutrinos from the $W$ and \tauhad decays, has a maximum around 20~\GeV\ and a significant tail up to about 80~\GeV.

Previous measurements at hadron colliders of \mbox{$W$ boson}
production with the
subsequent decay $W\to\tau\nu_\tau$ based on $p\bar{p}$ collisions
were reported by the UA1 collaboration~\cite{ref:WtaunuUA1} at
center-of-mass energies of $\sqrt{s}$ = 546~\GeV~and \mbox{$\sqrt{s}$
  = 630~\GeV} and by the CDF and D0
collaborations~\cite{ref:WtaunuCDF,ref:WtaunuD0} at a center-of-mass energy of 
\mbox{$\sqrt{s}$ = 1.8~TeV}.

In this Letter, we describe the measurement of this process with $\sqrt{s}=7$~TeV $pp$ collision data, which were recorded with the \mbox{ATLAS} experiment at the LHC. 

\section{The ATLAS detector}
\label{sec:detector}
The ATLAS detector is described in Ref.~\cite{ATLASDetector}. The cylindrical 
coordinate system is defined with polar angles~$\theta$ relative to the beamline and azimuthal angles~$\phi$ in the plane transverse to the beam. Pseudorapidities $\eta$ are defined as
$\eta = - \ln \tan \frac{\theta}{2}$. Transverse momenta, \pT, are defined as the component of momentum perpedicular to the beamline. 
Distances are measured in the $\eta$-$\phi$ plane as $\Delta R =
\sqrt{\Delta\eta^2+\Delta\phi^2}$. 

Measurements of charged-particle trajectories and momenta are performed with silicon detectors in the pseudorapidity range $|\eta|<2.5$, and also by a straw-tube tracking chamber in the range $|\eta|<2.0$. Together, these systems form the inner tracking detector, which is contained in a 2~T magnetic field produced by a superconducting solenoid. 
These tracking detectors are surrounded by a finely segmented calorimeter system which provides three-dimensional reconstruction of particle showers up to \linebreak\mbox{$|\eta| < 4.9$}. The electromagnetic calorimeter uses liquid argon as the active material and comprises separate barrel (\mbox{$|\eta| < 1.5$}), end-cap ($1.4 < |\eta| < 3.2$) and forward ($3.2 < |\eta| < 4.9$) components. The hadron calorimeter is based on scintillating tiles in the central region ($|\eta| < 1.7$). It is extended up to $|\eta| = 4.9$ by end-caps and forward calorimeters which use liquid argon.
The muon spectrometer measures the deflection of muon tracks in the field of three large superconducting toroidal magnets. It is instrumented with trigger and high-precision tracking chambers.

The trigger system 
consists of three levels. 
The first level is implemented as a hardware trigger, while the decision on the following levels is based on software event processing similar to the offline reconstruction.

\section{Data samples} 
\label{sec:data_mc}
The data used in this measurement were recorded in proton-proton collisions at a center-of-mass energy of \linebreak \mbox{$\sqrt{s}=7$~TeV} during the 2010 LHC run. The integrated luminosity of the data sample, considering only data-taking periods where all relevant detector subsystems were fully operational, is \lumi~\cite{lumipaper,lumi}.
The data were collected using triggers combining the two main signatures of \linebreak  \mbox{$W\to\tauhad\nu_\tau$} decays, namely the presence of a hadronically decaying $\tau$ lepton and missing transverse energy. 

Processes producing $W$ or $Z$ bosons that subsequently decay into
electrons or muons constitute important backgrounds to this
measurement if the lepton from the decay or an accompanying jet is
misidentified as a hadronically decaying $\tau$ lepton. Here, the
missing transverse energy signature arises from a $W$ decay neutrino
or the misreconstruction of jets or of other objects in the
event. Also, $W\to\tau\nu_\tau$ decays with the $\tau$ decaying
leptonically are considered as a background. Incompletely
reconstructed \Ztotautau and \ttbar decays can also enter the signal sample. The number of background events from these electroweak processes is referred to as \NEW in the following.

The production of $W$ and $Z$ bosons in association with jets is simulated with the PYTHIA~\cite{Pythia} generator with the modified LO parton distribution function (PDF) MRSTLO* \cite{ref:mrstlo} and normalized to the NNLO cross section;  \ttbar
processes are generated with MC@NLO~\cite{MCatNLO}, where parton showers and hadronization are simulated with HERWIG~\cite{Herwig} and the underlying event with JIMMY~\cite{JIMMY}. The TAUOLA \cite{TAUOLA} and PHOTOS~\cite{PHOTOS} programs are used to model the decay of $\tau$ leptons and the QED radiation of photons, respectively.

All simulated samples include multiple proton-proton interactions (pile-up) produced with PYTHIA using the ATLAS MC10 tune~\cite{pileup1}. Those samples are passed through a full detector simulation based on GEANT4 \cite{Geant4,ATLASSIM}. The simulated events are re-weighted so that the distribution of the number of reconstructed primary vertices per bunch crossing matches the data.

Due to their large production cross sections, QCD processes provide a significant background if quark/gluon jets (QCD jets) are misidentified as hadronic $\tau$ decays and a significant amount of \MET is measured, mainly due to incomplete reconstruction. The number of QCD background events \NQCD is estimated directly from data. 

\section{Object reconstruction}
\label{sec:objects}
Electron candidates, which together with muons are relevant for the electroweak background,
 are reconstructed from a cluster in the electromagnetic calorimeter matched to a track in the inner tracking detector. The cluster must have a shower profile consistent with an electromagnetic shower~\cite{ATL-PHYS-PUB-2011-006}. 
Muon candidates are reconstructed by combining tracks in the muon spectrometer 
with tracks in the inner tracking detector~\cite{ATLAS-CONF-2011-063}. 

Jets are reconstructed with the anti-$k_t$ algorithm~\cite{AntiKT} 
with a radius parameter $R=0.4$. The jet energies are calibrated~\cite{ATLASJETEnergyScale} using a $\pt$- and $\eta$-dependent calibration scheme, corrected for losses in dead material and outside the jet cone \cite{LocalHadronCalib}.
All jets considered in this analysis are required to have a transverse momentum above $20$~GeV 
and a pseudorapidity in the range $|\eta|<4.5$.

Reconstructed jets within $|\eta|<2.5$ provide the starting point (seed) for the reconstruction of  
hadronic $\tau$ decays.
The direction of a \tauhad\ candidate is taken directly from the corresponding seed jet. The energy is calibrated by applying a dedicated correction extracted from Monte Carlo to the sum of energies of the cells that form the clusters of the seed jet~\cite{ATLAS-CONF-2011-077}. Therefore, the energy of the \tauhad refers to the visible decay products. The transverse momentum is calculated as $\pt=E\sin\theta$, i.e.\  \tauhad\ candidates are treated as massless.
Good-quality tracks are associated with a \tauhad\ candidate if they are found within $\Delta R<0.2$ around the seed jet axis. At least one track must be associated to the candidate.

The \tauhad identification~\cite{ATLAS-CONF-2011-077} is based on eight observables: The invariant mass of the $\tau$ decay products is calculated separately using the associated tracks and the associated clusters. The fact that the $\tau$ decay products are typically more collimated than QCD jets is quantified by calculating the transverse momentum-weighted radius from tracks and the energy-weighted radius from electromagnetic energy information. The fraction of transverse energy within \mbox{$\Delta R <$ 0.1} of the \tauhad seed direction is used as well. 
Further discrimination is provided by the fraction of the transverse
\tauhad momentum carried by the highest-\pt track and the fraction of
transverse energy deposited in the electromagnetic calorimeter, for
which higher values are expected in case of hadronic $\tau$ decays
compared to QCD jets. For \tauhad candidates with more than one
associated track, the $\tau$ lifetime is also exploited by measuring the decay length significance of the associated secondary vertex in the transverse plane.
The single most discriminating of these quantities is the energy-weighted radius 
\begin{equation}
R_{\mathrm{EM}} =  \frac{\sum_{i}^{\Delta R_i< 0.4}E_{\mathrm{T},i}^{\mathrm{EM}}\Delta R_i}{\sum_{i}^{\Delta R_i< 0.4}E_{\mathrm{T},i}^{\mathrm{EM}}},
\end{equation}
where $i$ iterates over cells in the first three layers of the electromagnetic calorimeter associated with the \tauhad candidate, 
$\Delta R_i$ is defined relative to the \tauhad  seed axis, 
and $E_{\mathrm{T},i}^{\mathrm{EM}}$ is the cell transverse energy.

These eight variables are combined in a boosted decision tree discriminator (BDT)~\cite{boosting}, which provides an output value between 0 (background-like) and 1 (signal-like) with a continuous gradient of signal and background efficiency.
This discriminator was optimized using a combination of \Wtaunu and
\Ztautau Monte Carlo samples for the signal. The background was modeled from dijet events selected from data.
For \tauhad\ transverse momenta (\tauhadpT) above 20~\GeV, the efficiency of the \tauhad identification at the tighter working point of the BDT identification considered for this measurement is about 30\% with a jet rejection factor of 100 for \tauhad candidates with one track, while for candidates with three tracks it is about 35\% with a rejection factor of 300~\cite{ATLAS-CONF-2011-077}.
Additional requirements on the calorimeter and tracking properties of \tauhad candidates are used to discriminate against electrons and muons.

The missing transverse energy in the event, $\MET$, is reconstructed as $\sqrt{(\METx)^2+(\METy)^2}$, where $(\METx, \METy)$  is the vector sum of all calorimeter energy clusters 
in the region $|\eta|<4.5$, corrected for identified
muons~\cite{ref:perf}. With good approximation, the resolution of \MET
components is proportional to $a \times \sqrt{\SumET}$, where the
scaling factor $a$ depends on both the detector and reconstruction performance and $\SumET$ is calculated from all calorimeter energy clusters. The factor $a$ is about 0.5 $\sqrt{\mathrm{GeV}}$ for minimum bias events~\cite{ATLAS-CONF-MET-Calibration-7TeV}.

In order to reject events with large reconstructed  \MET due to fluctuations in the energy measurement, we define the significance of \MET as:
\begin{equation}
\METSIG = \frac{\MET [\mathrm{GeV}]}{0.5\sqrt{\mathrm{GeV}}\sqrt{(\SumET [\mathrm{GeV}])}}.
\end{equation}
\METSIG is found to provide better discrimination between the signal and the background from QCD jets than a simple \MET requirement. 
 
\begin{table*}[]
\centering
\begin{tabular}{lcccc}
\hline
\hline
 & A & B & C & D\\
\hline
\hline
$N^i$ (Data)&   2335&   4796&   1577&   27636\\
$N_{\mathrm{sig}}^i$ (\Wtaunu)&        1811  $\pm$ 25&      683   $\pm$ 16&      269   $\pm$ 8&       93    $\pm$ 5\\
$N_{\mathrm{\EW}}^i$&     284   $\pm$ 7&       118   $\pm$ 4&       388   $\pm$ 9&       90    $\pm$ 4\\
$c_{i}$& &      0.38    $\pm$ 0.01&      0.31    $\pm$ 0.01&      0.087    $\pm$ 0.003\\
\hline
$N_{\mathrm{QCD}}^i$&      127     $\pm$ 8&         3953    $\pm$ 75&        885    $\pm$ 45 &       27444   $\pm$ 166 \\\hline
\hline
\end{tabular}
\caption{Estimated sample compositions and $c_i$ factors (as defined in Equation \ref{eq:abcd:ci}) in the signal region A and control regions B, C, and D defined in the text.}
\label{tab:results:ABCD}
\end{table*}

\section{Event selection}
\label{sec:selection}
Events are selected using triggers based on the presence of a \tauhad jet and \MET. In the earlier part of the 2010 data taking, corresponding to an integrated luminosity of \mbox{11 \ipb}, a loosely identified \tauhad candidate with \mbox{$\tauhadpT >12$ \GeV}~(as reconstructed at the trigger level) in combination with $\MET > 20$ \GeV~was required. In the second part of the period \mbox{(24 \ipb)}, a tighter \tauhad identification and higher thresholds of 16~\GeV~and 22~\GeV~had to be used for \tauhadpT and \MET, respectively, due to the increasing luminosity.
The signal efficiencies of these two triggers with respect to the offline selection are estimated from the simulation to be (81.3$\pm$0.8)\% and (62.7$\pm$0.7)\%, respectively.

Events satisfying the trigger selection are required to have at least one reconstructed vertex that is formed by three or more tracks with \pT$>$ 150 MeV. Further selection requirements based on calorimeter information are applied to reject non-collision events and events containing jets that were incompletely reconstructed or significantly affected by electronic noise in the calorimeters. 
 
The calorimeter has a lower resolution for jets in the barrel-endcap transition regions. In order to ensure a uniform \MET resolution, events are rejected if a jet or a \tauhad candidate with $1.3<|\eta|<1.7$ is found.
In events where the \MET is found to be collinear to one of the jets, the reconstructed \MET is likely to originate from an incomplete reconstruction of this jet. Therefore, a minimum separation $|\Delta\phi(\mathrm{jet},\MET)| > 0.5$ rad is required.

In order to suppress backgrounds from other leptonic $W$ and $Z$ decays, events containing identified electrons or muons with  $\pt > 15 \GeV$ are rejected.
The \mbox{highest-\pt} identified \tauhad candidate in the event is considered for further analysis and required to be in the pseudorapidity range  $|\eta|<2.5$ and to have \mbox{$20 < \tauhadpT < 60$ \GeV}.  
A minimum \MET of 30~\GeV~is required and events are rejected if $\METSIG < 6$. 

\section{Background estimation}
\label{sec:back}

\begin{figure}[h!]
\centering
\includegraphics[width= 0.4\textwidth]{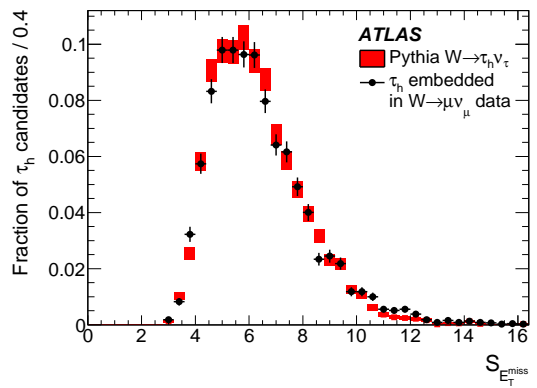}\label{fig:abcd:ComparisonPythiaMCembedded_DeltaPhi}
\label{fig:abcd:ComparisonPythiaMCembedded_METSignificance:METignificance_withoutTriggerCutmain}
\caption{Distribution of \METSIG for the \tauhad-embedded \Wmunu data sample (points) and simulated $W\to\tauhad\nu_\tau$ events (histogram), including statistical uncertainties.}
\label{fig:embedding}
\end{figure}

The number of expected events from signal and electroweak background processes is obtained from simulation. This is justified by the good agreement between data and simulation observed in the ATLAS $W$ cross section measurements~\cite{ATLAS-CONF-2011-041,WZnewpaper} through decays into electrons or muons. It is further validated using a high-purity data sample of $W\to\mu\nu_\mu$ events, in which the muon is removed and replaced by a simulated \tauhad lepton. Thus, only the $\tau$ decay and the corresponding detector response are taken from simulation while the underlying $W$ kinematics and all the other properties of the event are obtained from the $W\to\mu\nu_\mu$ events selected in data. Figure~\ref{fig:embedding} compares 
the distribution of \METSIG
for the \tauhad-embedded data sample with simulated $W\to\tauhad\nu_\tau$ events. A good agreement is observed within the statistical uncertainties, which adds further confidence in the electroweak background event model provided by the simulated event samples used in this analysis. 

The background contribution from QCD jet production, for which the cross section is large and the selection efficiency is low, cannot be reliably modeled using simulated events alone and is thus estimated from data.
In addition to the signal-dominated data set defined by the selection described in Section~\ref{sec:selection}, three background control regions are defined by inverting the requirements on the \METSIG and/or the \tauhad identification (ID), resulting in the following four samples:
\begin{itemize} 
\item Region A: $\METSIG > 6.0$ and \tauhad candidates satisfying the signal  \tauhad ID requirements described in Section \ref{sec:objects};
\item Region B: $\METSIG < 4.5$ and \tauhad candidates satisfying the signal-region \tauhad ID requirements; 
\item Region C: $\METSIG > 6.0$ and \tauhad candidates satisfying a looser \tauhad-ID but failing the signal-region \tauhad ID requirements;
\item Region D: $\METSIG < 4.5$ and \tauhad candidates satisfying a looser \tauhad-ID but failing the signal-region \tauhad ID requirements.
\end{itemize}
Here, the looser \tauhad-ID region is defined by selecting \tauhad candidates with a lower value of the BDT output.
 
After ensuring that the shape of the \METSIG distribution for the QCD background is independent of the \tauhad-ID requirement and assuming that the signal and electroweak background contributions in the three control regions are negligible, an estimate for the number of QCD background events in the signal region A is provided by
\begin{equation}
N\mathrm{_{QCD}^{A}} = N\mathrm{^{B}} N\mathrm{^{C}}/N\mathrm{^{D}},
\label{eq:abcd:NA_QCD}
\end{equation}
where $N^{i}$ represents the number of observed events in region $i$.

In order to take into account the residual signal and \EW~background contamination in the control regions\linebreak$i={\mathrm{B,C,D}}$, the number of selected events, ${N}^{i}$, needs to be replaced in Equation \ref{eq:abcd:NA_QCD} by 
$N^{i} - c_{i}(N\mathrm{^{A}}-N\mathrm{^{A}_{QCD}})$, 
where 
\begin{equation}
c_i = \frac{N^{i}\mathrm{_{sig}}+N^{i}\mathrm{_{\EW}}}{N\mathrm{_{sig}^{A}}+N\mathrm{_{\EW}^{A}}}
\label{eq:abcd:ci}
\end{equation} 
is the ratio of simulated signal and EW background events in the control region $i$ and the signal region. 
Therefore Equation \ref{eq:abcd:NA_QCD} becomes:
\begin{equation}
\hspace*{-0.88cm}N\mathrm{_{QCD}^{A}}\!\! =\!\! \frac{[N\mathrm{^{B}}-c\mathrm{_{B}}(N\mathrm{^{A}}-N\mathrm{^{A}_\mathrm{QCD}})][N\mathrm{^{C}}-c\mathrm{_{C}}(N\mathrm{^{A}}-N\mathrm{^{A}_{QCD}})]}{N\mathrm{^{D}}-c\mathrm{_{D}}(N\mathrm{^{A}}-N\mathrm{^{A}_{QCD}})}.
\nonumber
\label{eq:ABCD_corrected}
\end{equation}
The statistical error on $N\mathrm{^{A}_{QCD}}$ includes both the
uncertainty on the calculation of the $c_i$ coefficients, due to the
Monte Carlo statistics, and the statistical uncertainty of the data in the four regions. 
The resulting estimates of the sample compositions are summarized in Table~\ref{tab:results:ABCD}. 

\begin{figure}[b!]
\centering
\subfigure[]{\includegraphics[width= 0.4\textwidth]{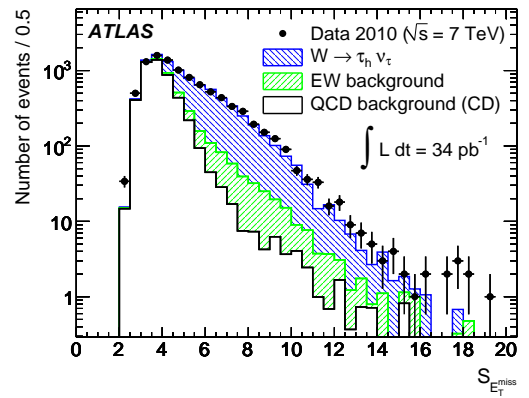}\label{fig:results:etmissig_AB_QCDCD}}\\
\subfigure[]{\includegraphics[width= 0.4\textwidth]{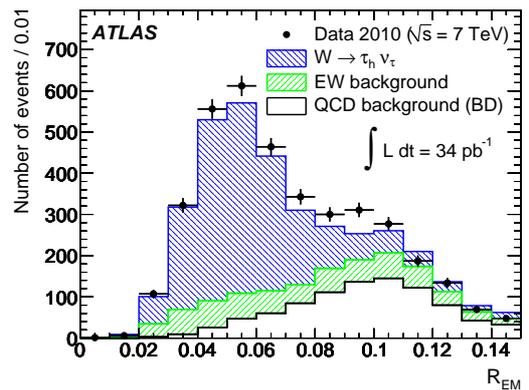}\label{fig:results:MET_rem_AC_QCDBD_log}
\label{fig:results:rem_et_AD_QCDC}}
\caption{
(a)\METSIG distribution in the combined region AB, extended over the full \METSIG range. 
The QCD background shape has been extracted from regions CD. 
Monte Carlo signal and \EW~background in regions AB are also shown; (b) the \tauhad identification variable
$R_{\mathrm{EM}}$ in the combined region AC. The QCD background shape has been extracted from regions BD. Monte Carlo signal and \EW~background in regions AC are also shown.
}
\label{fig:results:METSIG_AB}
\end{figure}

\begin{figure*}[t]
\centering
\subfigure[]{\includegraphics[width= 0.3\textwidth]{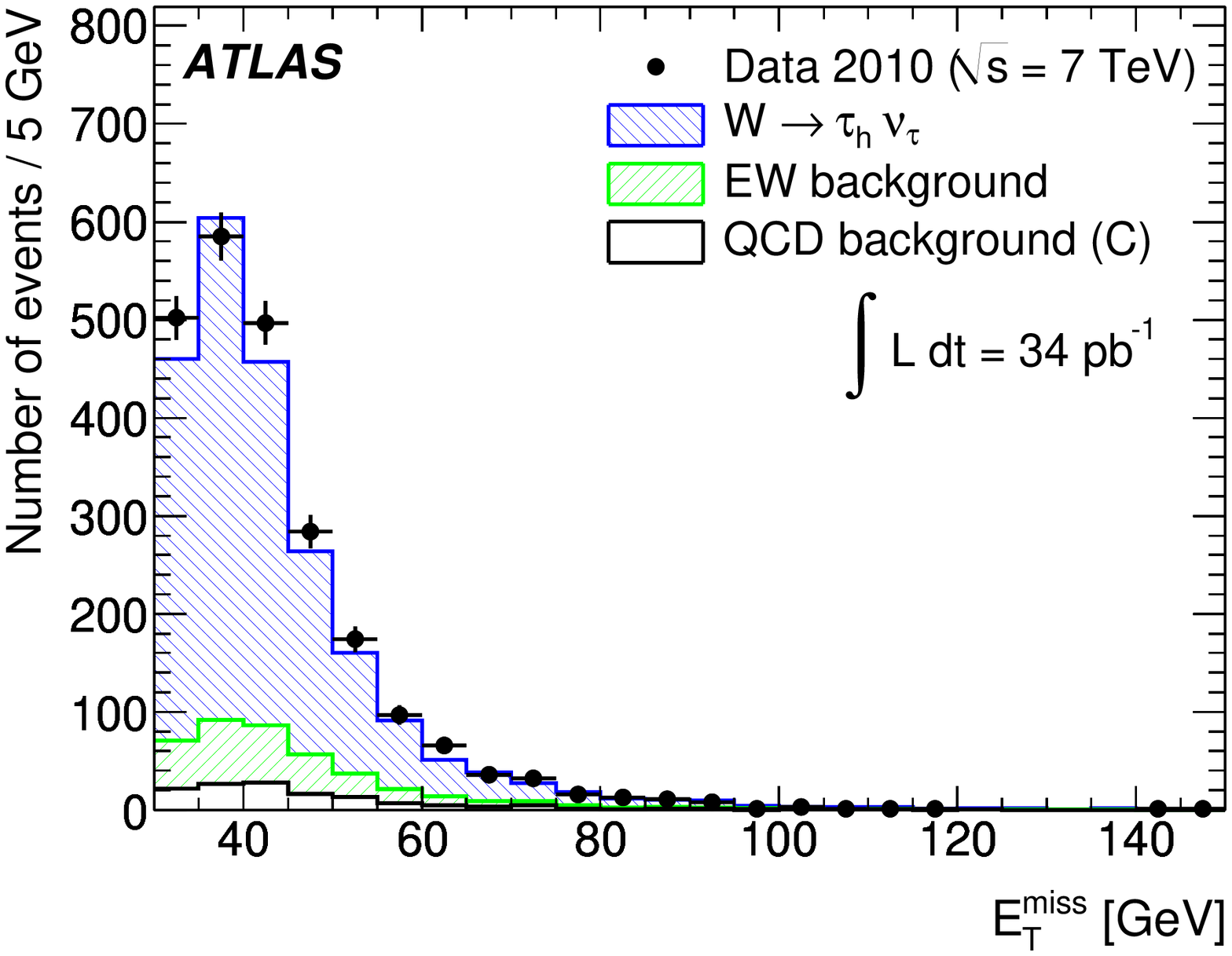}
\label{fig:results:MET_et_A_QCDC}}
\subfigure[]{\includegraphics[width= 0.3\textwidth]{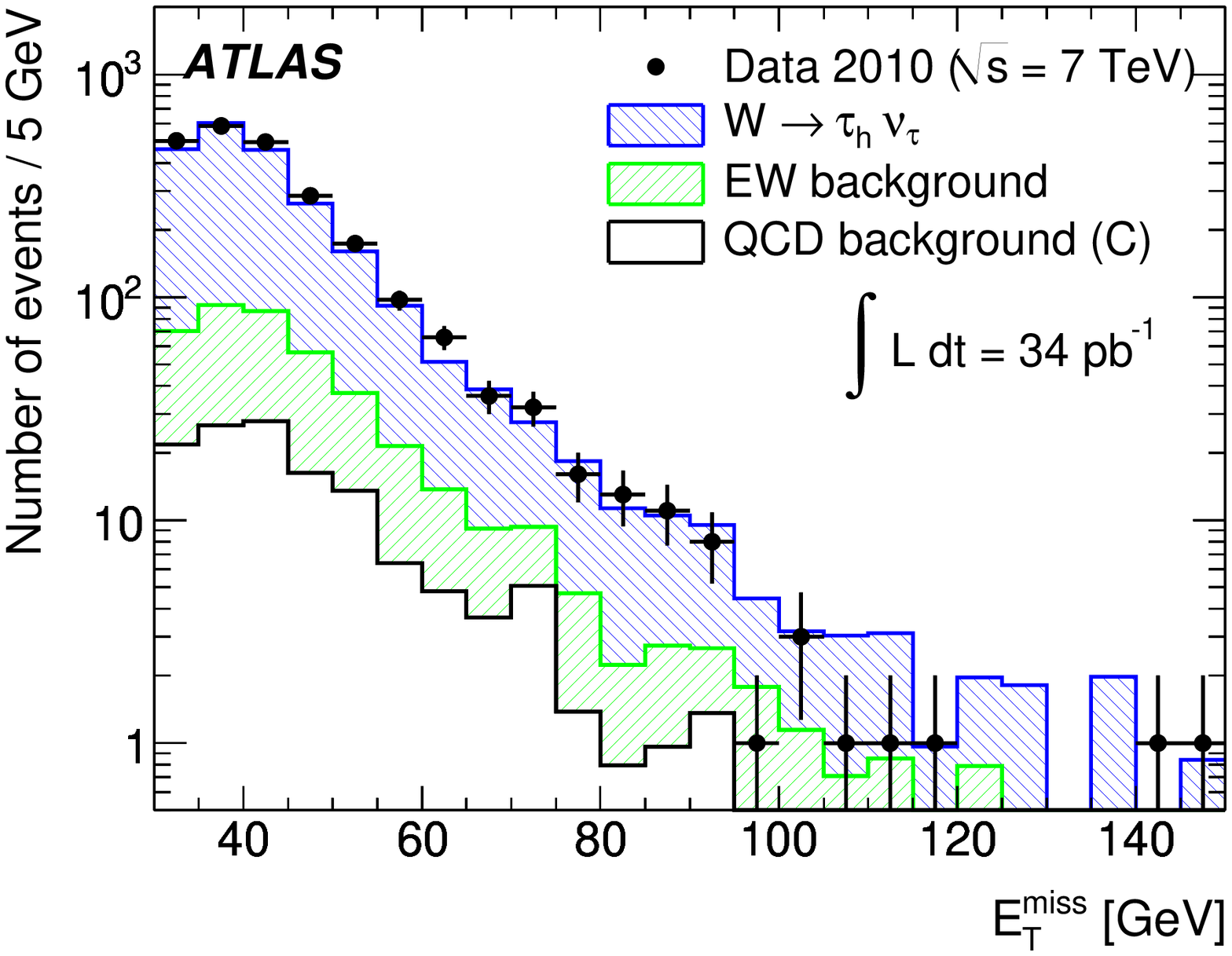}
\label{fig:results:MET_et_A_QCDC_log}}
\subfigure[]{\includegraphics[width= 0.3\textwidth]{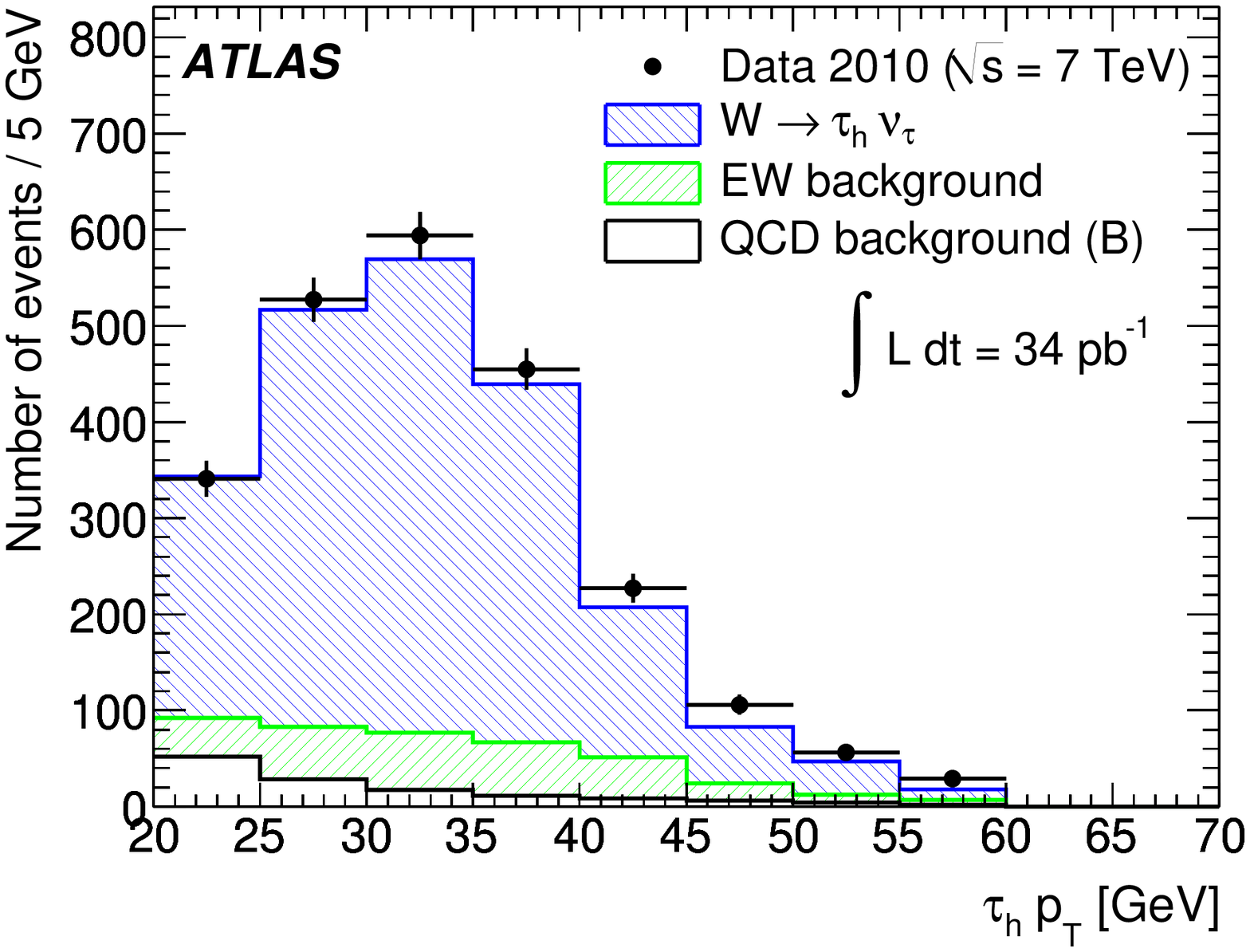}
\label{fig:results:tau_Et_A_QCDB}}
\subfigure[]{\includegraphics[width= 0.3\textwidth]{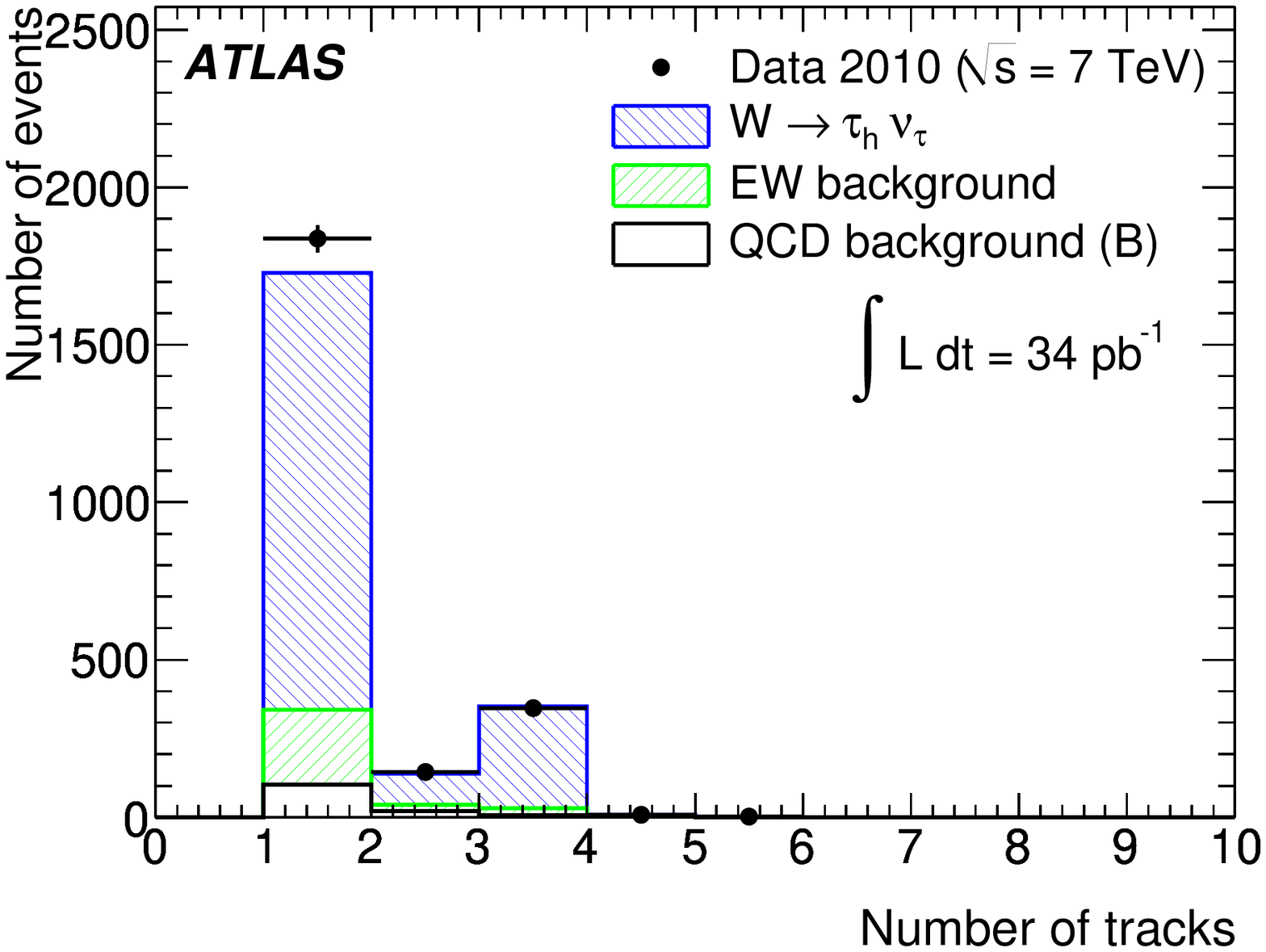}
\label{fig:results:tau_ntrack_A_QCDB}}
\subfigure[]{\includegraphics[width= 0.3\textwidth]{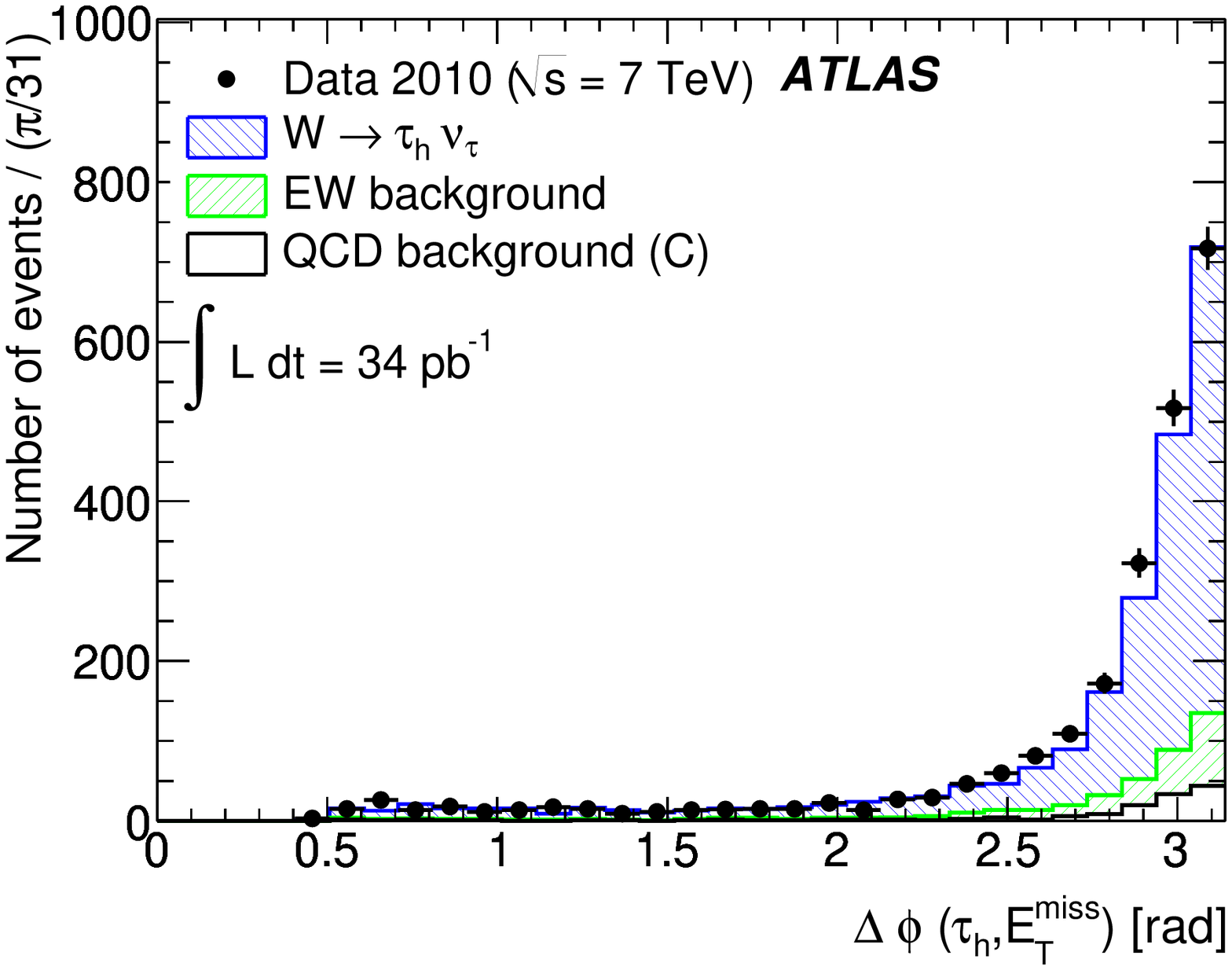}
\label{fig:results:DeltaPhi_A_QCDC}}
\subfigure[]{\includegraphics[width= 0.3\textwidth]{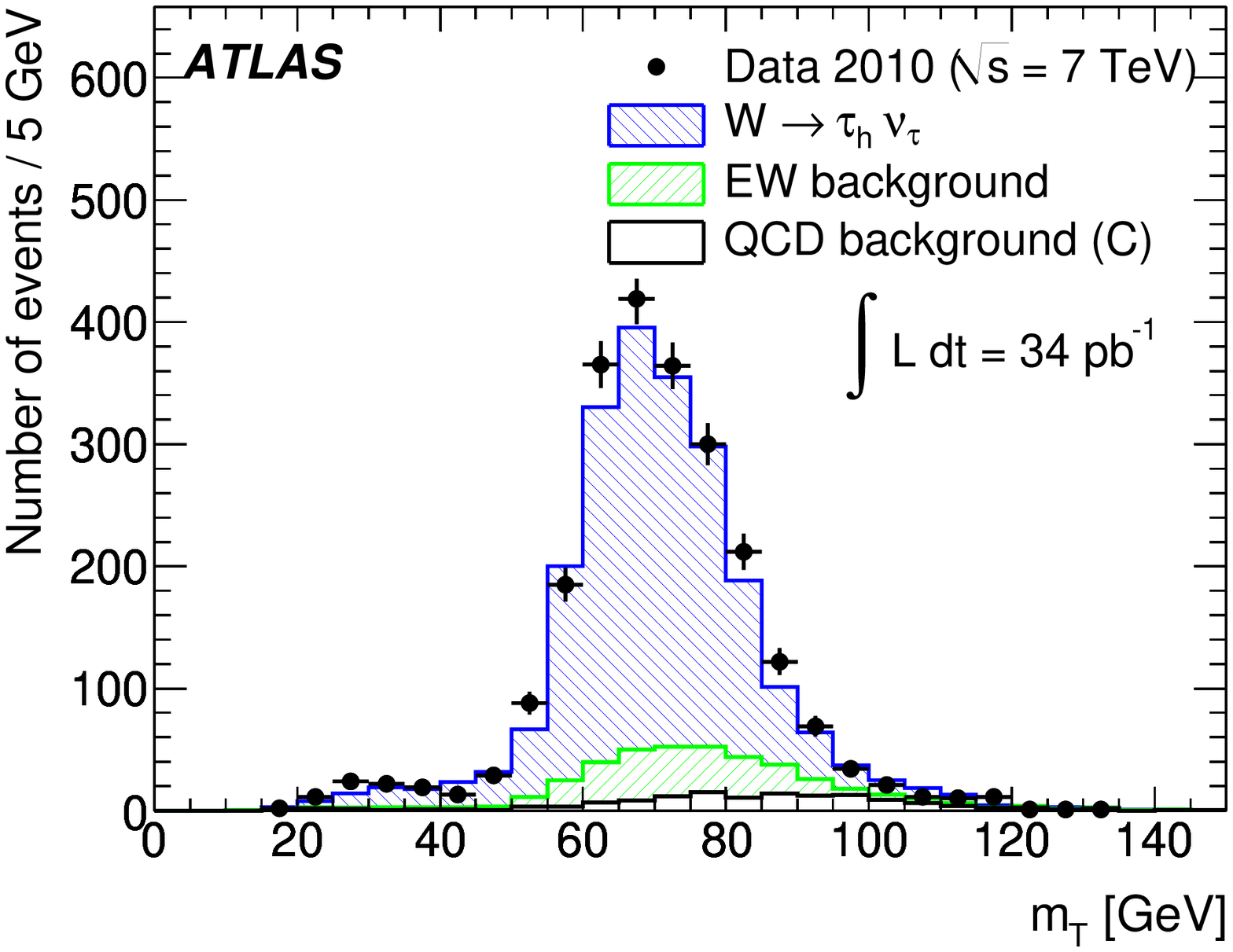}
\label{fig:results:MT_A_QCDC}}
\caption{
(a) Distribution of  missing transverse energy
in signal region A on a linear scale. The QCD background shape has been extracted from control region C.
(b) Same distribution on a logarithmic scale. 
(c) Transverse momentum and (d) number of  tracks of \tauhad 
candidates in signal region A. The QCD background shape has been extracted from control region B.
(e) Distribution of $\Delta\phi(\tauhad,\MET)$ and (f) transverse mass m$_{\mathrm{T}}$ in signal region A. The QCD background shape has been extracted from control region C.
The expectation from Monte Carlo signal and \EW~background in region A are also shown. 
}
\label{fig:results:Resultdistributions}
\end{figure*}

The quality of the description of the selected data by the background
models can be judged from \mbox{Figures~\ref{fig:results:METSIG_AB}
  and
  \ref{fig:results:Resultdistributions}}, where data and the
background estimates (EW and QCD) are shown. Figure~\ref{fig:results:etmissig_AB_QCDCD}
shows the distribution of \METSIG in regions A and B, extended
over the full \METSIG range, for all events passing the selection
criteria except for the \METSIG requirement. In
Figure~\ref{fig:results:rem_et_AD_QCDC} the distribution of
$R_{\mathrm{EM}}$ is shown. In this case events passing the selection
criteria but considering \tauhad candidates identified by the loose
and the tight selections in regions A and C are shown. The agreement between data and Monte Carlo
expectation confirms the results obtained by the data-driven
background estimation. In Figure~\ref{fig:results:Resultdistributions}
the distribution of \MET, the \tauhadpT spectrum, the number of tracks
associated to the \tauhad candidate, the distribution of
$\Delta\phi(\tauhad,\MET)$ and the transverse mass, $m_{\mathrm T} = \sqrt{2\cdot p_{\mathrm T}^{\mathrm{\tauhad}}\cdot \MET \cdot \left(1- \cos{\Delta \phi\left(\tauhad, \MET\right)}\right)}$, in the selected signal region A are shown, illustrating the characteristic properties of \Wtaunu decays.
In all the distributions reasonable agreement is observed between the data and Monte Carlo prediction.

\section{Cross section measurement}
\label{sec:xsecintro}
The fiducial cross section is measured in a phase space region given by the geometrical acceptance of the detector and by the kinematic selection of the analysis (as described in Section \ref{sec:selection}). This region is defined based on the decay products from a simulated hadronic $\tau$ decay and corresponds to the criteria presented in Table \ref{tab:xsecintro:acc}.
\begin{table}[h!]
\centering
\begin{tabular}{c}
\hline
\hline
$20 \GeV < \pt^{\tau,\mathrm{vis}} < 60 \GeV$ \\
$|\eta^{\tau,\mathrm{vis}}| < 2.5$, excluding $1.3 < |\eta^{\tau,\mathrm{vis}}| < 1.7$ \\
$(\sum p^\nu)_T > 30 \GeV$ \\
$|\Delta\phi(p^{\tau,\mathrm{vis}},\sum p^\nu)| > 0.5$ \\
\hline
\hline
\end{tabular}
\caption{Definition of the acceptance region.}
\label{tab:xsecintro:acc}
\end{table}

Here, the visible $\tau$ momentum $\pt^{\tau,\mathrm{vis}}$ and pseudorapidity $\eta^{\tau,\mathrm{vis}}$ are calculated from the sum of the four-vectors of the decay products from the simulated hadronic $\tau$ decay, except for the neutrinos. This momentum also includes photons radiated both from the $\tau$ lepton and from the decay products themselves, considering only photons within $\Delta R < 0.4$ with respect to the \tauhad. The minimum \MET requirement translates into a cut on the transverse component of the sum of the simulated neutrino four-vectors $(\sum p^\nu)_T$.

The fiducial cross section, including the branching ratio $BR(W\rightarrow\tauhad\nu_\tau)$, is computed as
\begin{equation}
\sigmafid 
 = \frac{N_{\mathrm{obs}} - N_{\mathrm{bkg}}}{C_W \cal{L}},
\label{eq:sigfid}
\end{equation}
where $N_{\mathrm{obs}}$ is the number of observed events in data, $N_{\mathrm{bkg}}$ is the number of estimated (QCD and \EW) background events (signal region A in Table \ref{tab:results:ABCD}), and $\cal{L}$ is the integrated luminosity. 
 $C_W$ is the correction factor that takes into account the efficiency of trigger,  \tauhad reconstruction and identification and the efficiency of all selection cuts within the acceptance:
\begin{equation}
C_W = \frac{N_{\mathrm{reco,~all~cuts}}}{N_{\mathrm{gen,~kin/geom}}},
\label{eq:xsecintro:cw}
\end{equation}
where $N_{\mathrm{reco,~all~cuts}}$ is the number of fully simulated signal events passing the reconstruction, trigger and the selection cuts of the analysis and $N_{\mathrm{gen,~kin/geom}}$ is the number of simulated signal events within the fiducial region defined above.  

With the kinematic and geometrical signal acceptance 
\begin{equation}
A_W = \frac{N_{\mathrm{gen,~kin/geom}}}{N_{\mathrm{gen,~all}}},
\label{eq:xsecintro:aw}
\end{equation}
where $N_{\mathrm{gen,~all}}$ is the total number of simulated signal events while $N_{\mathrm{gen,~kin/geom}}$ is the denominator of $C_W$, the total cross section
\begin{equation}
\sigmatot = \sigmafid /A_W = \frac{N_{\mathrm{obs}} - N_{\mathrm{bkg}}}{A_W C_W \cal{L}}
\label{eq:sigtot}
\end{equation}
can be obtained.
$A_W$ and $C_W$ are determined using a  PYTHIA Monte Carlo signal sample 
described in Section \ref{sec:data_mc}.
The fiducial acceptance is found to be $A_W = \AW \pm~\AWstat \mathrm{~(MC~stat)}$ and the correction factor $C_W = \CW \pm \CWstat \mathrm{~(MC~stat)}$.

The measured fiducial cross section of the \Wtaunu decay is
$\sigmafid =  \xsecfid \pm \xsecfidstat \mathrm{~(stat)} \mathrm{~nb}$
and the total cross section is found to be
$\sigmatot= \xsec \pm \xsecstat \mathrm{~(stat)}\mathrm{~nb}$.

Several alternative analyses are performed to confirm these results. For example,  
the BDT \tauhad ID is replaced by a simpler identification based on cuts on three of the ID variables only~\cite{ATLAS-CONF-2011-077}. 
Also, in order to study the influence of pile-up on the result, the signal selection is restricted to events with only one reconstructed primary vertex. 
In both cases consistent results are found.

\section{Systematic uncertainties}
\label{sec:syspaper}

Table~\ref{tab:syspaper:sysuncert} summarizes the systematic uncertainties.
The main sources 
are discussed in the following.

\begin{table*}[]
\begin{center}
\begin{tabular}{lcccc}
\hline\hline
& $\frac{\delta C_W}{C_W}$ 						        & $\frac{\delta \NEW}{\NEW}$ 
                       & $\frac{\delta \NQCD}{\NQCD}$& $\frac{\delta \sigmafid}{\sigmafid}$	\\ 
\hline\hline
Trigger efficiency	&6.1\%  & 6.1\%	&-  & 7.0\%    \\
Energy scale		&6.7\%  & 8.7\% &-     &8.0\%\\
\tauhad ID efficiency	&9.6\%  & 4.1\% &-     &10.3\%\\
Jet \tauhad misidentification	& -     & 7.2\% &-   & 1.1\%  \\
Electron \tauhad misidentification	& -     & 4.5\% &- &0.7\%     \\
Pile-up reweighting 	&1.4\%  & 1.2\%	&-     &1.6\%\\
Electron reconstruction/identification		
                        & -     & 1.2\% &-    &0.2\% \\
Muon reconstruction	& -     & 0.3\% &-     &0.04\%\\
Underlying event modeling		&1.3\%  & 1.1\% &-     &1.5\%\\
Cross section	        &-      & 4.5\%	&-    & 0.7\%\\
\hline
QCD estimation: Stability/correlation   
                        & -     & -     &2.7\% &0.2\%\\
QCD estimation: Sig./\EW~contamination  
                        & -     & -     &2.1\% &0.1\%\\
\hline
Monte Carlo statistics           &    1.4\%   &   2.4\%    &   6.0\%   & 1.5\%\\
\hline \hline
Total systematic uncertainty	& 13.4\%& 15.2\%&6.9\%&15.1\%\\
\hline\hline
\end{tabular}
\normalsize
\end{center}
\caption{Summary table for systematic uncertainties. For the systematic uncertainty on the fiducial cross section measurement, correlations between the systematics affecting $C_W$  and $N\mathrm{_{EW}}$ have been taken into account.}
\label{tab:syspaper:sysuncert}
\end{table*}

\paragraph{Monte Carlo predictions} \ \\
The trigger efficiency is determined in Monte Carlo for the
combined \MET and \tauhad triggers used in the two data periods. The differences between the
measured trigger responses of the two trigger components in data and Monte Carlo are used to determine the systematic uncertainty.
A pure and unbiased sample enriched with \Wtaunu events is obtained in data by applying an independent \tauhad (\MET) trigger and selected cuts of the event selection like the BDT \tauhad ID.  The corresponding \MET (\tauhad) trigger part is applied to this sample and the response of this trigger is compared to the response in Monte Carlo. The observed differences are integrated over the offline \tauhadpT and \MET range used for the cross section measurement. 
The total systematic uncertainty after the combination of the different trigger parts is 6.1\%.

The signal and background acceptance depends on the energy scale of the clusters used in the computation of \MET and \METSIG and the energy scale of the calibrated \tauhad candidates. Based on the current knowledge of the calibration the uncertainty due to cluster energy within the detector region $|\eta|<3.2$ is at most 10\%\ for \pt of \mbox{500 \MeV} and within 3\% at high \pt~\cite{ATLAS-CONF-2011-080}. In the forward region $|\eta|>3.2$ it is estimated to be 10\%.
The effect on \MET and \METSIG has been evaluated by scaling all clusters in the event according to these uncertainties and recalculating \MET and \SumET. 
At the same time, the \tauhad energy scale has been varied according to its uncertainty~\cite{ATLAS-CONF-2011-077}. This uncertainty depends on the number of tracks associated to the \tauhad candidate, its \pt and the \mbox{$\eta$ region} in which it was reconstructed, and ranges from 2.5\% to 10\%.
In addition, the sensitivity of the signal and background efficiency to the \MET resolution has been investigated~\cite{ATLAS-CONF-MET-Calibration-7TeV}.
Consequently, the yield of signal and \EW~background varies within 6.7\% and 8.7\%, respectively. 

The identification and reconstruction efficiency of \tauhad candidates 
was studied with Monte Carlo \Wtaunu and \Ztautau samples and was
found to vary with different simulation conditions such as
different underlying event models, detector geometry, hadronic shower
modeling and noise thresholds for calorimeter cells in the cluster
reconstruction. In Ref.~\cite{ATLAS-CONF-2011-077}, these uncertainties are evaluated as a function of \tauhadpT, separately for candidates with one or multiple tracks and low or high multiplicity of primary vertices in the event. The corresponding changes in the signal and \EW~background efficiencies are found to be 9.6\% and 4.1\%, respectively.

The probability of a jet or electron to be misidentified as a \tauhad candidate has been evaluated in data and compared with the expectation from Monte Carlo. 
The rate of jets that are misidentified as \tauhad candidates was calculated using a selection of W$\rightarrow \ell \nu_\ell +$jets events (with $\ell = e,\mu$) and measuring the fraction of reconstructed candidates that are found by the \tauhad identification. The difference of this misidentification rate in Monte Carlo compared to that in data is 30\% and this was applied as a systematic uncertainty to the fraction of events mimicked by a jet. The overall uncertainty on the \EW~background is 7.2\%. 
The misidentification probability of electrons as \tauhad candidates has been determined with a ``tag-and-probe'' method using \Zee events where the \tauhad identification and \tauhad electron veto is applied to one of the electrons. The difference between the misidentification probability in data and Monte Carlo as a function of \eta~has been applied as a systematic uncertainty to \tauhad candidates mimicked by an electron. It amounts to 4.5\% for the total \EW~background.

Other sources of systematic uncertainty have been evaluated and were found to have only small effects on the resulting cross section measurement, 
for example the procedure to include pile-up effects, 
the uncertainty on the lepton selection efficiency entering via the veto of electrons and muons and the influence of the underlying event modeling on \MET quantities.
The uncertainties on the cross sections used for the \EW~background are taken from \mbox{ATLAS} measurements, when available, or theoretical \linebreak NNLO calculations, and lie between 3 and 9.7\%~\cite{Aad:2010yt,ref:xsec:top,ref:FEZW,ref:MSTW2008}.
The uncertainty on the integrated luminosity is 3.4\%~\cite{lumipaper,lumi}.

\paragraph{QCD background estimation} \ \\ 
Two different sources of systematic uncertainty arising from the method of estimating the QCD background events from data have been studied.
The stability of the method and the small correlation of the two variables (\tauhad ID and $\METSIG$) used to define the control regions have been tested by varying the $\METSIG$ threshold.
The systematic uncertainty due to the correction for signal and \EW~background contamination in the control regions was obtained by varying the fraction of these events in the regions within the combined systematic and statistical uncertainties on the Monte Carlo predictions discussed above. 
The total uncertainty on the QCD background estimation is 3.4\%.

\paragraph{Acceptance} \ \\
The theoretical uncertainty on the geometric and kinematic acceptance factor $A_W$ is dominated by the limited knowledge of the proton PDFs and the modeling of \linebreak \mbox{$W$ boson} production at the LHC.

The uncertainty resulting from the choice of the PDF set is evaluated by comparing the acceptance obtained with different PDF sets (the default MRST LO*, CTEQ6.6 and HERAPDF 1.0 \cite{ref:pdf}) and within one PDF set by re-weighting the default sample to the different error eigenvectors available for the CTEQ6.6 NLO PDF~\cite{CTEQ66}. The uncertainty is 1.6 \% and 1.0\%, respectively, which 
combines to 1.9\%. 

The uncertainty on the modeling of $W$ production was evaluated by comparing the default sample acceptance to that obtained from an MC@NLO sample where the parton shower is modeled by HERWIG. 
The difference in acceptance is found to be smaller than 0.5\%.
 
\section{Results}
\label{sec:xsec}

The results of the analysis relevant to the cross section measurement are summarized in Table~\ref{tab:xsec:results}.
\begin{table}[h!]
\centering
\begin{tabular}{ll}
\hline
\hline
$N_{\mathrm{obs}}$ 	& \NA \\
\NQCD  & \NAQCDal $\pm$ \NAQCDsys \\
\NEW	& \NAEW $\pm$ \NAEWsys \\
$A_{W}$ 	& \AW$\pm$ \AWsys \\
$C_{W}$ 	& \CW $\pm$ \CWsys\\
\hline
\hline
\end{tabular}
\caption{Resulting numbers for the cross section calculation. The errors include statistical and systematic uncertainties here.}
\label{tab:xsec:results}
\end{table} 
Within the acceptance region defined in Table~\ref{tab:xsecintro:acc} they translate into a fiducial cross section \sigmafid of: 
\begin{eqnarray*}
\xsecfid \pm \xsecfidstat~\mathrm{(stat)} \pm \xsecfidsyst~\mathrm{(syst)} \pm \xsecfidlumi~\mathrm{(lumi)} \mathrm{~nb}
\label{eq:xsec:fid}
\end{eqnarray*}
and a total cross section \sigmatot of:
\begin{eqnarray*}
\xsec \pm \xsecstat~\mathrm{(stat)} \pm \xsecsyst~\mathrm{(syst)} \pm \xseclumi~\mathrm{(lumi)} \mathrm{~nb}.
\label{eq:xsec:totbr}
\end{eqnarray*}
After correcting the cross section for the hadronic $\tau$ decay branching ratio $BR(\tau\rightarrow h\nu_\tau)$ = 0.6479 $\pm$ 0.0007~\cite{ref:xsec:PDG} this yields the following inclusive cross section \sigmatotfullw:
\begin{eqnarray*}
\xsectot \pm \xsectotstat~\mathrm{(stat)} 
\pm \xsectotsys~\mathrm{(syst)} \pm \xsectotlumi~\mathrm{(lumi)} \mathrm{~nb}.
\label{eq:xsec:tot}
\end{eqnarray*}

\begin{figure}[t!]
\centering
\hspace*{-0.5cm}
 \includegraphics[width= 0.48\textwidth]{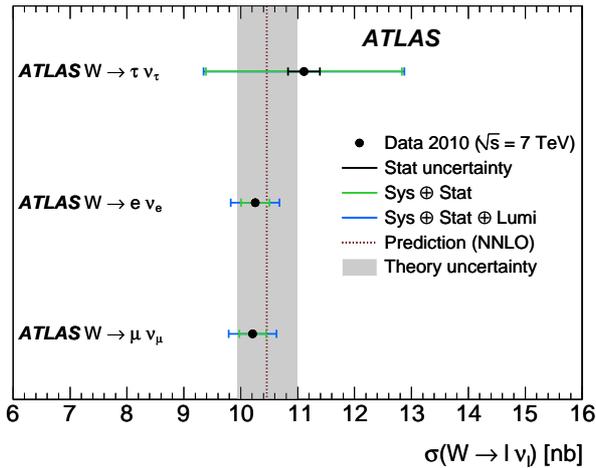}
 \caption{
Cross sections for the different $\ensuremath{W\to\ell\nu_\ell}\xspace$ channels measured in ATLAS with 2010 data (points). Systematic, luminosity and statistical uncertainties are added in quadrature. The theoretical NNLO expectation is also shown (dashed line), together with its uncertainty (filled area).}
 \label{fig:xsec:comparexsec}
 \end{figure}

The measured cross section is in good agreement with the theoretical NNLO cross section $10.46 \pm 0.52$~nb~\cite{ref:FEZW,ref:MSTW2008,Aad:2010yt} and the ATLAS measurements of the \Wenu and \mbox{\Wmunu} cross sections~\cite{ATLAS-CONF-2011-041,WZnewpaper}. 
The comparison of the cross section measurements for the different lepton final states and the theoretical expectation is shown in Figure~\ref{fig:xsec:comparexsec}. This is the first $W\to\tau\nu_\tau$ cross section measurement performed at the LHC.
\clearpage

\section{Acknowledgements}

We thank CERN for the very successful operation of the LHC, as well as the
support staff from our institutions without whom ATLAS could not be
operated efficiently.

We acknowledge the support of ANPCyT, Argentina; YerPhI, Armenia; ARC,
Australia; BMWF, Austria; \newline ANAS, Azerbaijan; SSTC, Belarus; CNPq and FAPESP,
Brazil; NSERC, NRC and CFI, Canada; CERN; CONICYT, Chile; CAS, MOST and
NSFC, China; COLCIENCIAS, Colombia; MSMT CR, MPO CR and VSC CR, Czech
Republic; DNRF, DNSRC and Lundbeck Foundation, Denmark; ARTEMIS, European
Union; IN2P3-CNRS, CEA-DSM/IRFU, France; GNAS, Georgia; BMBF, DFG, HGF, MPG
and AvH Foundation, Germany; GSRT, Greece; ISF, MINERVA, GIF, DIP and
Benoziyo Center, Israel; INFN, Italy; MEXT and JSPS, Japan; CNRST, Morocco;
FOM and NWO, Netherlands; RCN, Norway; MNiSW, Poland; GRICES and FCT,
Portugal; MERYS (MECTS), Romania; MES of Russia and ROSATOM, Russian
Federation; JINR; MSTD, Serbia; MSSR, Slovakia; ARRS and MVZT, Slovenia;
DST/NRF, South Africa; MICINN, Spain; SRC and Wallenberg Foundation,
Sweden; SER, SNSF and Cantons of Bern and Geneva, Switzerland; NSC, Taiwan;
TAEK, Turkey; STFC, the Royal Society and Leverhulme Trust, United Kingdom;
DOE and NSF, United States of America.

The crucial computing support from all WLCG partners is acknowledged
gratefully, in particular from CERN and the ATLAS Tier-1 facilities at
TRIUMF (Canada), NDGF (Denmark, Norway, Sweden), CC-IN2P3 (France),
KIT/GridKA (Germany), INFN-CNAF (Italy), \newline NL-T1 (Netherlands), PIC (Spain),
ASGC (Taiwan), RAL (UK) and BNL (USA) and in the Tier-2 facilities
worldwide.

\bibliographystyle{atlasnote}
\bibliography{WtoTauNu2010}

\newpage
 
\clearpage
\onecolumn
% ATLAS Collaboration author list for 07-JUL-2011
% Data extracted on 22-Jul-2011 for paperid 105
%\documentclass[11pt]{article}
%\usepackage{a4wide}\begin{document}
\begin{flushleft}
{\Large The ATLAS Collaboration}

\bigskip

G.~Aad$^{\rm 48}$,
B.~Abbott$^{\rm 111}$,
J.~Abdallah$^{\rm 11}$,
A.A.~Abdelalim$^{\rm 49}$,
A.~Abdesselam$^{\rm 118}$,
O.~Abdinov$^{\rm 10}$,
B.~Abi$^{\rm 112}$,
M.~Abolins$^{\rm 88}$,
H.~Abramowicz$^{\rm 153}$,
H.~Abreu$^{\rm 115}$
% ...
\bigskip

% ...
$^{10}$ Institute of Physics, Azerbaijan Academy of Sciences, Baku, Azerbaijan\\
$^{11}$ Institut de F\'isica d'Altes Energies and Departament de F\'isica de la Universitat Aut\`onoma  de Barcelona and ICREA, Barcelona, Spain\\
% ...
$^{48}$ Fakult\"{a}t f\"{u}r Mathematik und Physik, Albert-Ludwigs-Universit\"{a}t, Freiburg i.Br., Germany\\
$^{49}$ Section de Physique, Universit\'e de Gen\`eve, Geneva, Switzerland\\
% ...
$^{88}$ Department of Physics and Astronomy, Michigan State University, East Lansing MI, United States of America\\
% ...
$^{111}$ Homer L. Dodge Department of Physics and Astronomy, University of Oklahoma, Norman OK, United States of America\\
$^{112}$ Department of Physics, Oklahoma State University, Stillwater OK, United States of America\\
% ...
$^{115}$ LAL, Univ. Paris-Sud and CNRS/IN2P3, Orsay, France\\
% ...
$^{118}$ Department of Physics, Oxford University, Oxford, United Kingdom\\
% ...
$^{153}$ Raymond and Beverly Sackler School of Physics and Astronomy, Tel Aviv University, Tel Aviv, Israel\\
% ...
\end{flushleft}

%\end{document}

\end{document}